\documentclass[a4paper,10pt]{article}

\usepackage{jcappub}
\usepackage[utf8x]{inputenc}
\usepackage{multicol}
\usepackage[font=footnotesize,labelfont=bf]{caption}
\usepackage{float}
\usepackage{tensor}

\title{Cosmological study of intersecting s-branes solutions in M-theory including dark matter}
\author[a]{ Agudelo, J. A.}
\author[b]{Idarraga, A. }
\author[c]{ and Arcos, H. I. }
\affiliation[a]{\small Universidade Federal de Mato Grosso,\\ Instituto de F\'{i}sica\\ 
Av. Fernando Corr\^{e}a da Costa 2367, Cuiab\'{a} - MT - Brasil}
\affiliation[b]{\small Universidade Federal do ABC,\\ Centro de Matem\'{a}tica, Computa\c{c}\~{a}o e Cogni\c{c}\~{a}o
\\ Av. dos Estados 5001, Santo Andr\'{e} - SP - Brasil}
\affiliation[c]{\small Universidad Tecnol\'{o}gica de Pereira,\\ Facultad de Ciencias B\'{a}sicas\\ 
Vereda La Julita, Pereira - RIS - Colombia}
\emailAdd{jaar@fisica.ufmt.br}
\emailAdd{a.lopez@ufabc.edu.br}
\emailAdd{hiarcosv@utp.edu.co}
\abstract{A cosmological analysis considering the inclusion of dark energy and dark matter cosmic components in the context of some particular cases for intersecting s-brane solutions is presented and discussed. Direct solution of $d$-dim field equations and dimensional reduction processes, are implemented for the pure dark energy and dark energy with dark matter cosmological scenarios, respectively. In the first case, explicit expressions and evolution for $d$-dim scale factors are founded and studied. Secondly, a low energy effective $4$-dim model is obtained analytically and their resulting field equations are solved numerically. The role of internal space geometry is always a central element of our analysis. In some intersecting cases and under certain considerations, a suitable late-time cosmic acceleration description is founded and an adequate behavior for scale factor $a(t)$ and realistic values for dark energy $\Omega_{DE}$ and dark matter $\Omega_{DM}$ relative energy densities are obtained as well. As a result, we do specifically obtain late-time cosmic acceleration in one scenario where our universe makes part  either of an SM2$\bot$SM2 or an  SM2$\bot$SM5 intersection. This could give rise to further possible configurations where late-time cosmic acceleration is present.}
\keywords{cosmology with extra dimensions, dark energy theory, dark matter theory, cosmological simulations}
\toccontinuoustrue
\begin{document}
%
%
\maketitle
\flushbottom
%
%
\begin{abstract}

\end{abstract}
%
%
%
%
\section{Introduction}
Many recent observational data suggest that our universe is currently under an accelerated 
expansion phase \cite{Weinberg:2013}-\cite{Riess:1998}, so the problem of dark energy and, even more, its 
explanation in the context of fundamental theories such as M/string theory, have gained considerable 
importance during last years. Although, it is possible to build cosmological models 
with the desired features by suitable modifications of Einstein's gravity \cite{Turner:2007}\cite{Tsujikawa:2011}, 
it is imperative to deduce some cosmological effects of current cosmic state from super-gravity theories 
\cite{Gutperle:2003}.\newline
Today, we know that in order to derive such effects from compactifications in super-gravity theories, it is 
necessary to consider the compact internal space like a manifold with time-dependent volume, since otherwise, 
a called \emph{no-go theorem} appears excluding the possibility of getting accelerated cosmic expansion under these 
models, because they naturally satisfy a strong energy condition (SEC) \cite{Maldacena:2000}\cite{Teo:2005}.\newline
It has found that this \emph{no-go theorem} can be avoided if it is included time-dependence for internal space, 
obtaining a solution for a vacuum universe under hyperbolic compactification \cite{Townsend:2003}, which corresponds to 
one particular case of a general solution for an arbitrary number of orthogonally intersecting S-branes, into a super-gravity model including dark energy by a dilatonic-scalar coupling \cite{Ohta:2003}\cite{Ohta:2003-2}. \newline
In this context, new analytic solutions have taken place, in the case of SM2 and SD2 branes, describing accelerating cosmologies from compactifications, only for plane and hyperbolic internal space \cite{Chen:2003}\cite{Roy:2003}. However, it was also showed that, under certain considerations and for certain time interval, spherical case works as well \cite{Ohta:2003S}.\newline
We shall give continuity to this kind of developments, studying the possibility of describing accelerating and effective $4$-dim cosmologies from $d$-dim super-gravity, but constructing some intersecting s-branes solutions using the harmonic rules discussed in \cite{Deger:2002} and including the dark matter component in a phenomenological setting, as was proposed in \cite{Gutperle:2003}. Our main purpose consists in showing that these intersecting s-branes scenarios can describe the current cosmic acceleration according to observational data.\newline
The document is structured as follows. In Section 2, essential details about the model and the two different implemented methods are presented. The harmonic intersecting rules, the particular cases of study and the basis of dimensional reduction process are presented as well. In section 3, cosmological properties of the particular considered solutions are analyzed and discussed. In section 4, the conclusions and perspectives are condensed.
%
%
\section{The Model and Methodology}
We start with the action \cite{Ohta:2003}
\begin{equation}\label{2.1}
 S_{d}= \kappa_{d}\int d^{d}x\sqrt{-g}\left[R_{d}
-\frac{1}{2}\left(\partial\phi\right)^{2}
  -\sum_{A=1}^{m}\frac{1}{2n_{A}!}e^{a_{A}\phi}F^{2}_{n_{A}}\right],
\end{equation}
which describes the $d$-dimensional ($d=4+n$) gravity coupled to the dilaton and $m$ different $n_A$-form field strengths, 
representing the electromagnetic interaction of each S-brane. In the above equation, $\kappa_{d}\equiv 1/(16\pi G_{d})$.\newline
In this paper, we study cosmic acceleration from this model with two different but complementary methods: the first one 
consists of solving directly $d$-dim equations of motion, focusing our attention in the scale factors' time evolution for 
a completely filled dark energy universe without other matter-energy form presence. The second one, corresponds to a dimensional 
reduction process by the calculation of Ricci components, considering that the entire $d$-dim space (and hence its 
associated metric tensor) can be decomposed in two sub-spaces, namely, an observable part or external and, a non-observable 
part or internal, which can be represented with a $4$-dim Lorentzian and $n$-dim semi-Riemannian manifolds, 
respectively. \newline
Subsequently, it is possible to build a $4$-dim action from \eqref{2.1} and to make the corresponding cosmological 
analysis, studying like in the first case, scale factors' time evolution and, additionally, complementing the model with 
dark matter contribution in a phenomenological way, introducing a dark energy-momentum tensor.
%
%
\subsection{Direct Solution of d-dim Field Equations}
The action \eqref{2.1} has been solved by N. Ohta in \cite{Ohta:2003}, taking the metric 
\begin{equation}\label{2.2}
  ds_{d}^{2}=-e^{2u_{0}}d\xi^{2}+\sum_{\alpha=1}^{p}e^{2u_{\alpha}}d{y_{\alpha}}^{2}+
e^{2B}d\Sigma^{2}_{n,\sigma},
\end{equation}
where $u_{o}$, $u_{\alpha}$ and $B$ are functions only of $\xi$ and chosen to satisfy a gauge condition 
given by
\begin{equation}\nonumber
 -u_{0}+\sum_{\alpha=1}^{p}{u_{\alpha}}+nB=0,
\end{equation}
which is used to simplify the equations of motion. Finding the corresponding field equations to \eqref{2.1} for the 
above metric, it is possible to find expressions for $u_{o}$, $u_{\alpha}$, $B$, so that the solution 
takes the form
\begin{eqnarray}\label{2.3} \nonumber
 ds_{d}^{2}&=&\prod_{A}\left[\cosh\widetilde{c_A}(\xi-\xi_A)\right]^{2\frac{q_{A}+1}
{\Delta_A}}\Bigl\{e^{2ng(\xi)+2c_{0}\xi+c'_{0}}\left[-d\xi^{2}+e^{-2(n-1)g(\xi)}d\Sigma^{2}_{n,\sigma}\right]\\
&&+\sum_{\alpha=1}^{p}\prod_{A}\left[\cosh\widetilde{c}_A(\xi-\xi_A)\right]^{-2\frac{\gamma_{A}^{(\alpha)}}
{\Delta_A}}e^{2\widetilde{c_{\alpha}}\xi+2c'_{\alpha}}dy_{\alpha}^{2}\Bigr\},
\end{eqnarray}
which represents a general solution for an arbitrary number $m$ of orthogonally intersecting S-branes. 
One can show, for M-theory, that the above general solution defines a simple set of rules, 
through which, it is possible to build a solution for each intersection case of S-branes consistent with 
the rules discussed in \cite{Deger:2002} (deduction of these rules is included for completeness in appendix \ref{A}). In order to clarify the analysis, we shall present 
its principal ideas as follows.
%
%
\subsection{Intersecting S-branes Solutions Rules in M-Theory}
The starting point consists of introducing several constants, three by each S-brane $q_{i}$, $\xi_{i}$, $M_{i}$, 
corresponding to electric charge, a given instant of time  and a positive number, respectively. One takes 
$\Sigma_{n,\sigma}$, as the $n$-dim transverse hyper-surface , which corresponds to an unitary sphere, 
hyperbola or plane space, depending on the $\sigma$ value (-1, 0 and 1, respectively). For an $S_p$-brane, the $11$-dimensions 
can be parameterized by $(x_{1},...,x_{p+1},y_{1},...,y_{q}$, $\xi$, $\Sigma_{n,\sigma})$, where 
$x $ and $ y$ parameterize tangent and relative common directions 
and $\xi,\Sigma_{n,\sigma}$ denote the general transverse or internal space.\newline
Furthermore, for each brane, one defines the following function depending on $q_{i}$, $\xi_{i}$ y $M_{i}$
\begin{equation}\label{2.4}
    H_{i}=\left(\frac{q_{i}}{M_{i}}\right)^{2}\cosh^{2}\left[M_{i}(\xi-\xi_{i})\right].
\end{equation}
Additionally to $H_{i}$, we need other function characterizing the general transverse space, this is
\begin{equation}\label{2.5}
    G_{n,\sigma}=\left\{
           \begin{array}{ll}
             \frac{2n(n-1)}{M^{2}}\sinh^{2}[\sqrt{\frac{n-1}{2n}}M|\xi|], & \hbox{$\sigma=-1$;} \\
             \frac{2n(n-1)}{M^{2}}\cosh^{2}[\sqrt{\frac{n-1}{2n}}M\xi], & \hbox{$\sigma=+1$;} \\
            e^{-2M\sqrt{\frac{n-1}{2n}}\xi}, & \hbox{$\sigma=0$.}
           \end{array}
         \right.,
\end{equation}
with $M^{2}=\sum_{i}M_{i}^{2}$. In terms of these functions, the corresponding metric for a fixed number of possible intersections, i.e., their solution, can be obtained from the next simple rules.
\begin{itemize}
  \item General transverse space takes the form:
  \begin{equation}\label{2.6}
    G_{n,\sigma}^{-\frac{n}{(n-1)}}\left[-d\xi^{2}+G_{n,\sigma}d\Sigma_{n,\sigma}^{2}\right].
  \end{equation}
  \item We can find the metric form, multiplying brane and transverse directions by suitable powers of $H_i$-functions, 
such that 
      \begin{equation}\label{2.7}
        \mbox{$SM2-branes$}\left\{
          \begin{array}{ll}
            \mbox{brane directions}, & \hbox{$H_{i}^{-1/3}$;} \\[0.25cm]
            \mbox{transverse direction}, & \hbox{$H_{i}^{1/6}$.}
          \end{array}
        \right.,
      \end{equation}
      \begin{equation}\label{2.8}
        \mbox{$SM5-branes$}\left\{
          \begin{array}{ll}
            \mbox{brane directions}, & \hbox{$H_{j}^{-1/6}$;} \\[0.25cm]
            \mbox{transverse direction}, & \hbox{$H_{j}^{1/6}$.}
          \end{array}
        \right..
      \end{equation}
\end{itemize}
In other words, each $H_i$ function appears in the metric as $H^{1/2}$, multiplying transverse directions and there is a conformal general factor $H^{-1/3}$ for $SM2$-branes and $H^{-1/6}$ for $SM5$-branes. \newline
On the other hand, following these rules for three interesting cases, namely SM2-brane, SM2$\bot$SM2(0) and SM2$\bot$SM5(1), 
one can analyze their solutions cosmologically through comparison with the standard metric for super-gravity given by
\begin{equation}\label{2.9}
    ds^{2}=\delta^{-n}(\xi)\left[-S^{6}(\xi)d\xi^{2}+S^{2}(\xi)dx^{2}_{\alpha}\right]+
\delta^{2}(\xi)d\Sigma^{2}_{n,\sigma},
\end{equation}
obtaining expressions for the observable and non-observable scale factors $S(\xi)$ and $\delta(\xi)$ 
respectively, and then analyzing their time evolution.\footnote{Note that we have used here the same notation 
$\bot$ and $(n)$ to refer to an intersection and the dimension of the resulting S-brane respectively, like in all previous 
related and cited papers.} Additionally, if the substitution,
\begin{equation}\label{2.10}
    dt=S^{3}(\xi)d\xi,
\end{equation}
is imposed, we clearly have a FRLW universe with scale factor $S(\xi)$, which should satisfy the conditions
\begin{equation}\label{2.11}
    P(\xi)=\frac{dS}{dt}> 0   \qquad and \qquad Q(\xi)=\frac{d^{2}S}{dt^{2}}>0,
\end{equation}
in order to describe positive acceleration.
%
%
\subsubsection{SM2-brane} 
In the presence of a single SM2-brane, the $11$-dim space-time can be decomposed as (11=3+1+7), so that $n=7$ and
\begin{equation}\label{2.12}
   ds^{2}=H^{-1/3}\left[dx_{1}^{2}+dx_{2}^{2}+dx_{3}^{2}\right]+H^{1/6}G_{7,\sigma}^{-7/6}
   \left[-d\xi^{2}+G_{7,\sigma}d\Sigma_{7,\sigma}^{2}\right].
\end{equation}
Therefore, comparing with \eqref{2.9}, we have
\begin{eqnarray}\label{2.13}
    \nonumber \delta(\xi)=\left(\frac{H}{G_{7,\sigma}}\right)^{1/12}\\
    S(\xi)=H^{-1/6}\delta^{7/2}(\xi)=\left(\frac{H^{3/7}}{G_{7,\sigma}}\right)^{7/24}.
\end{eqnarray}
Thus, we find expressions that allow us to study the cosmic evolution in terms of observable and non-observable scale factors, 
where both depend on the specific form of $G_{n,\sigma}$ function according to \eqref{2.5}. Analyzing the 
hyperbolic case ($\sigma=-1$), we see that our results are in complete agreement with the expressions given by 
Gutperle et. al. in \cite{Gutperle:2003}, having the relationship between constants $q\equiv b$ and 
$\frac{q}{\sqrt{84}}\equiv\frac{b}{3}\sqrt{\frac{3}{28}}$, when $M=3$.
%
%
\subsubsection{\boldmath{$SM2\bot SM2(0)$}}\label{sec2.2.2}
For this case, we have $n=5$ and the space-time is decomposed as (11=1+2+2+1+5), so we get
\begin{align}\label{2.14}
   \nonumber  ds^{2}=& \,\left(H_{1}H_{2}\right)^{-1/3}[dx^{2}]+H_{1}^{-1/3}H_{2}^{1/6}[dy_{1}^{2}+dy_{2}^{2}]+
     H_{1}^{1/6}H_{2}^{-1/3}[dy_{3}^{2}+dy_{4}^{2}]\\
    & \,+\left(H_{1}H_{2}\right)^{1/6}G^{-5/4}_{5,\sigma}\left[-d\xi^{2}+G_{5,\sigma}d\Sigma^{2}_{5,\sigma}
\right].
\end{align}
For the sake of simplicity, we consider 
\begin{enumerate}
  \item Both $SM2-Branes$ have the same charge: $q_{1}=q_{2}=q$.
  \item Positive numbers $M_i$ are the same too: $M_{1}=M_{2}=m \Rightarrow 
M=\sqrt{M_{1}^{2}+M_{2}^{2}}=\sqrt{2}m$.
  \item Time instants $\xi_i$ which each brane assumes $q$ and $M$ previous point defined values, are equal to 
zero, this is, considering a time-scale shift: $\xi_{1}=\xi_{2}=0$.
\end{enumerate}
Applying these restrictions, is evident that as a consequence, we would have $H_{1}=H_{2}=H$ and, therefore, the metric 
\eqref{2.14} becomes
\begin{equation}\label{2.15}
    ds^{2}=H^{-2/3}[dx^{2}]+H^{-1/6}[dy_{1}^{2}+dy_{2}^{2}]+H^{-1/6}[dy_{3}^{2}+dy_{4}^{2}]+
H^{2/6}G^{-5/4}_{5,\sigma}\left[-d\xi^{2}+G_{5,\sigma}d\Sigma^{2}_{5,\sigma}\right].
\end{equation}
In order to get an uniform expanding universe, we propose the next re-definition of $x$ (the common coordinate) as
\begin{equation}
  \nonumber x \rightarrow H^{-1/4}\widetilde{x}, \hspace{1cm} \text{or} \hspace{1cm}
  \nonumber dx^{2} \rightarrow H^{-1/2}d{\widetilde{x}}^{2},
\end{equation}
representing a time variable measure along the common coordinate direction and letting us deduce expressions for scale factors analytically, again, by comparison with \eqref{2.9}. After this process, we get
\begin{equation}\label{2.16}
    ds^{2}=H^{-1/6}\left\{[dx^{2}]+\left[dy_{1}^{2}+dy_{2}^{2}\right]+\left[dy_{3}^{2}+dy_{4}^{2}\right]\right\}+
H^{2/6}G^{-5/4}_{5,\sigma}\left[-d\xi^{2}+G_{5,\sigma}d\Sigma^{2}_{5,\sigma}\right],
\end{equation}
and, finally
\begin{equation}\label{2.17}
  \delta(\xi)=H^{1/6}G^{-1/8}_{5,\sigma} \hspace{1cm}\text{and}\hspace{1cm}  S(\xi)=\delta^{5/2}(\xi)H^{-1/2}.
\end{equation}
If we consider now that $(q_1,M_1)\neq(q_1,M_2)$ which yields $H_1 \neq H_2$, we can see that, in addition for this case, we could propose the common coordinate re-definition as
\begin{equation}
  \nonumber dx \rightarrow H_{2}^{1/2}d\widetilde{x} \hspace{1cm} \text{or} \hspace{1cm} 
 \nonumber dx \rightarrow H_{1}^{1/2}d\widetilde{x},
\end{equation}
where the last assumptions would correspond to consider that spatial coordinates of the observable universe lie in just one of the S2-branes, such that the remaining ones will not be considered in our analysis. It is possible to show that for both re-definitions, scale factors are so similar, differing only in one exponent (-2), changing from $H_1$ to $H_2$ as the case may. If some restrictions are imposed to the constants $q_i$, $M_i$ prior to common coordinate re-definition, the above options would yield completely different results.
%
%
\subsubsection{SM2\boldmath{$\bot$}SM5(1)}
For this last case, $d$-dim space-time is decomposed as (11=2+1+4+1+3) and the metric takes the form
\begin{align}\label{2.18}
 \nonumber ds^{2}=&\,H_{1}^{-1/3}H_{2}^{-1/6}\left[dx_{1}^{2}+dx_{2}^{2}\right]+H_{1}^{-1/3}H_{2}^{1/3}\left[dy_{1}^{2}\right]+
H_{1}^{1/6}H_{2}^{-1/6}\left[dy_{2}^{2}+\ldots+dy_{5}^{2}\right]+\\
&\,H_{1}^{1/6}H_{2}^{1/3}G_{3,\sigma}^{-3/2}\left[-d\xi^{2}+G_{3,\sigma}d\Sigma_{3,\sigma}^{2}\right].
\end{align}
Now, we see that in order to describe an FRLW-type universe plus a compact internal space of dimension $n=3$, we must consider that our $4$-dim universe lies in $dx_{1}, dx_{2}$ and $dy_{1}$ coordinates, neglecting the 
$dy_{2}^{2}+\ldots+dy_{5}^{2}$ coordinates belonging to SM5-brane, obtaining
\begin{equation}\label{2.19}
 ds^{2}=H_{1}^{-1/3}H_{2}^{-1/6}\left[dx_{1}^{2}+dx_{2}^{2}\right]+H_{1}^{-1/3}H_{2}^{1/3}\left[dy_{1}^{2}\right]+
H_{1}^{1/6}H_{2}^{1/3}G_{3,\sigma}^{-3/2}\left[-d\xi^{2}+G_{3,\sigma}d\Sigma_{3,\sigma}^{2}\right].
\end{equation}
Let us propose, in this case, the next re-definition for $dy_{1}$, of the form
\begin{equation}
  \nonumber dy_{1} \rightarrow H_{2}^{-1/4}d\widetilde{y}_{1} \hspace{1cm} \text{or} \hspace{1cm}
  \nonumber dy_{1}^{2} \rightarrow H_{2}^{-1/2}d\widetilde{y}_{1}^{2},
\end{equation}
such that \eqref{2.19} transforms in
\begin{equation}\label{2.20}
 ds^{2}=H_{1}^{-1/3}H_{2}^{-1/6}\left[dx_{1}^{2}+dx_{2}^{2}+d\widetilde{y}_{1}^{2}\right]+
H_{1}^{1/6}H_{2}^{1/3}G_{3,\sigma}^{-3/2}\left[-d\xi^{2}+G_{3,\sigma}d\Sigma_{3,\sigma}^{2}\right].
\end{equation}
Therefore, we would have
\begin{equation}\label{2.21}
 \delta(\xi)=H_{1}^{1/12}H_{2}^{1/6}G^{-1/2}_{3,\sigma} \hspace{1cm}\text{and}\hspace{1cm}  
S(\xi)=\delta^{3/2}(\xi)H_{1}^{-1/6}H_{2}^{-1/12}.
\end{equation}
%
%
\subsection{Dimensional Reduction and Effective Potential}
This method is applied when it has been verified that cosmological behavior of scale factors in $d$-dim is suitable 
(last section), i. e., when they satisfy the conditions in \eqref{2.11}. It consists in a compactification that depends on the geometry of internal space considered as a compact manifold \cite{Carr:1986}. The result is a low 
energy model obtained from $d-dim$ M/string theory, coupled to a dilaton $\phi$, a radionic (scalar) field $\psi$ 
that characterizes the volume of internal space and, a scalar potential field which depends on the geometry of 
internal space and the field strength form \cite{Roy:2003}. See appendix \ref{B}. The resulting action in $4$-dim from \eqref{2.1} in 
$d$-dim is
\begin{equation}
   \nonumber S_{4}=\kappa_{(4)}\int{d^{4}x\sqrt{-g}\left({R-\frac{1}{2}\partial_{\mu}\phi\partial^{\mu}\phi
    -\frac{n(n+2)}{2}\partial_{\mu}\psi\partial^{\mu}\psi-V(\phi,\psi)}\right)},
\end{equation}
where
\begin{equation}
    V(\phi,\psi)=\frac{b^{2}}{2}\hspace{0.2cm}e^{-\frac{2(n-4)}{n+2}\phi-3n\psi}
    -\sigma n(n-1)e^{-(n+2)\psi}.\label{2.22}
\end{equation}
For solutions in M-theory, the dilatonic coupling is taken null. One can find the equations 
of motion from \eqref{2.22} applying the variational principle (with respect to the metric and  scalar fields). It is 
remarkable that the solutions of these equations following from the $4-dim$ Lagrangian are the same as 
the ones from $11-dim$. Since we are interested in describing cosmic evolution, considering both dark 
energy and dark matter, it is important to mention that currently the mechanisms for including dark matter at $d−dim$ level are 
not clear. However, in the phenomenological setting, one can introduce it by adding its energy density to the energy density of the scalar field in the corresponding equations of $4-dim$ effective model. 
Therefore, the equations of motion that 
take into account the main constituents of the universe and consider a spatially flat FRLW metric, are
\begin{align}\label{2.23}
    \ddot{\varphi}=& \,-3H\dot{\varphi}-V'\\
    H^2=& \,\frac{1}{3}\left(\rho_{DM}+\frac{1}{2}{\dot{\varphi}}^2+V(\varphi)\right)
\end{align}
where, $\varphi=\sqrt{\frac{63}{2}}\psi$ and $H=\frac{\dot{a}}{a}$. Then, we can solve numerically these equations and plot the time  evolution of the scale factor $a(t)$, the scalar field $\varphi(t)$, the relative dark energy density $\Omega_{DE}$, and the 
equation of state $\omega(z)$, where $z$ is the red-shift defined via the scale factor as $z=a^{-1}(t)-1$. Moreover $a(t)$ 
is taken to be $1$ at the present time, corresponding to $z=0$. In other words, we shall present this analysis 
process for the SM2-brane, SM2$\bot$SM2 and SM2$\bot$SM5 intersecting cases.
\begin{figure}[t]
\centering
\includegraphics{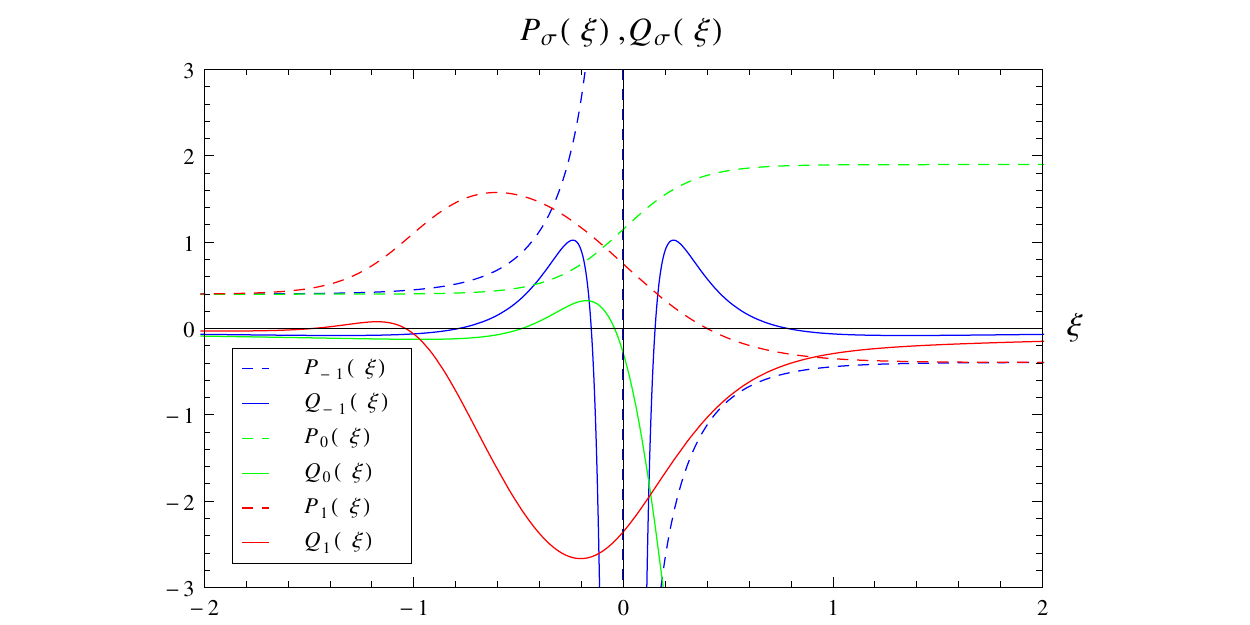}\\
\caption{The conditions to get acceleration \eqref{2.11}, where the solid and dashed 
lines correspond to $P(\xi)$ and  $Q(\xi)$, respectively. Blue, green and red colors correspond to $\sigma$ values of -1, 0 and 1. There is a 
discontinuity for blue lines, due to the universe evolves separately from $-\infty$ to 0 or from 0 to $\infty$ for hyperbolic case, but this 
problem does not appear in the other two cases where the universe evolves 
continuously from $-\infty$ to $\infty$.}\label{fig1}
\end{figure}
%
%
\section{Cosmological Analysis}
%
%
\begin{figure}[b]
(a)\hspace{-1cm}\includegraphics[scale=0.8]{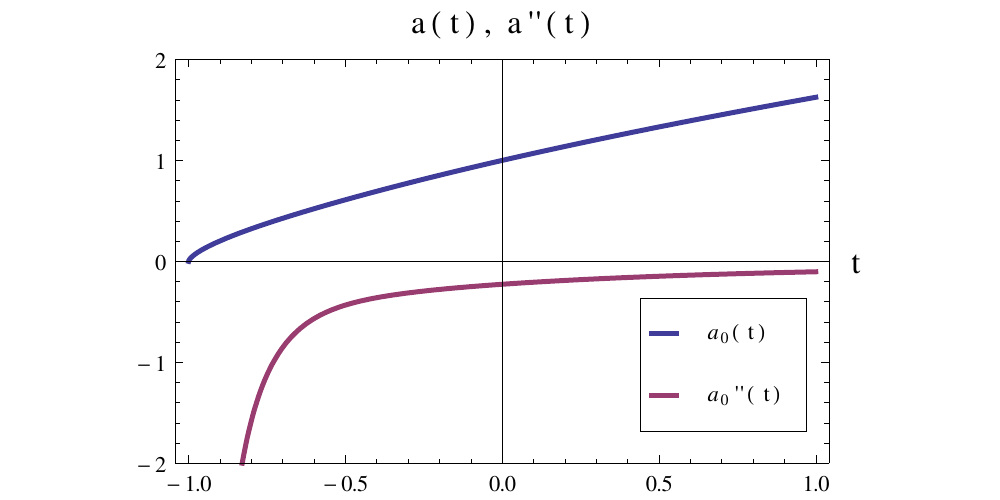}
(b)\hspace{-1cm}\includegraphics[scale=0.8]{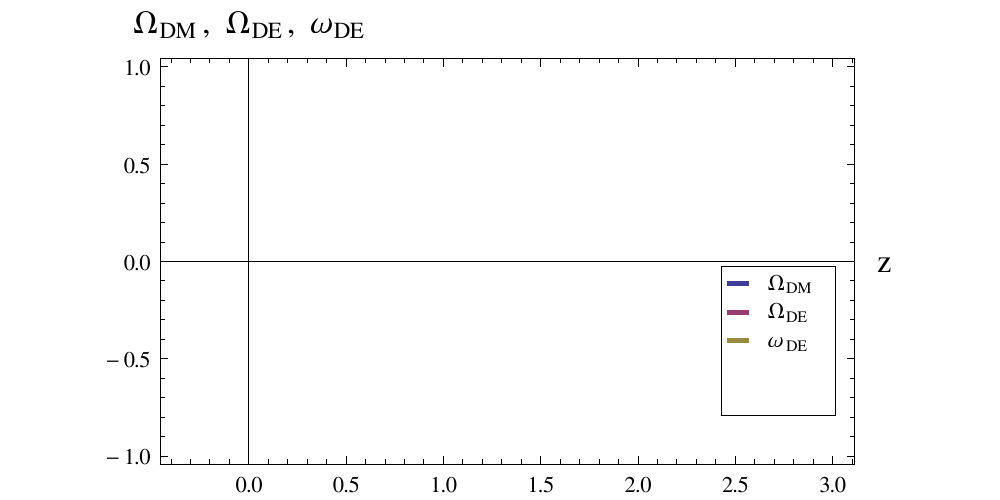}
\caption{In (a) we have plotted the scale factor $a(t)$ and its second 
derivative $a''(t)$. In (b) dark matter and dark energy relative densities $\Omega_{DM}$, $\Omega_{DE}$ and state equation parameter 
$\omega_{DE}$ have been plotted, showing their evolution as cosmological red shift $z=a^{-1}(t)-1$ functions, where $z=0$ corresponds to current time.}\label{fig2}
\end{figure}
\subsection{SM2-brane}
In figure \ref{fig1} we have plotted the conditions \eqref{2.11} for $d$-dim observable scale factor $S(\xi)$ and 
we see that is possible to describe positive acceleration for all $\sigma$ values for certain intervals of time. 
However, once you want to supplement the model with CDM phenomenologically, solving \eqref{2.23} numerically, the unique 
case that works according to observational data is $\sigma=-1$, as was shown previously in \cite{Gutperle:2003}. The 
reason lies in $4$-dim scale factor $a(t)$, relative dark energy density $\Omega_{DE}$ and state equation parameter 
$\omega_{DE}$ evolution for $\sigma=0$ case, and on the other hand, the effective potential behavior for $\sigma=1$ 
case, as follows.
\paragraph{Plane Case:} In figure \ref{fig2} (a), we see that despite $a(t)$ grows in time as one expects, its acceleration, although increases, is always negative tending to zero at infinity. This yields a decelerating universe which tends to a constant expansion. Additionally, the numerical values of $\Omega_{DE}\approx0.09$ and $\omega_{DE}\approx-0.75$ described in \ref{fig2} (b), are inconsistent with the current accepted values for dark matter, dark energy and the last restrictions for state equation parameter, 
which say that they must be $\Omega_{DM}\approx0.3$, $\Omega_{DE}\approx0.7$ and $\omega_{DE}<-0.78$, respectively. This particular behavior can be explained as a consequence of the addition of CDM, which would be responsible for the negative acceleration due its 
attractive gravitational force. Note that once you add CDM under these considerations, it becomes dominant yielding deceleration. 
\begin{figure}[t]
\centering
(a)\hspace{-1cm}\includegraphics[scale=0.8]{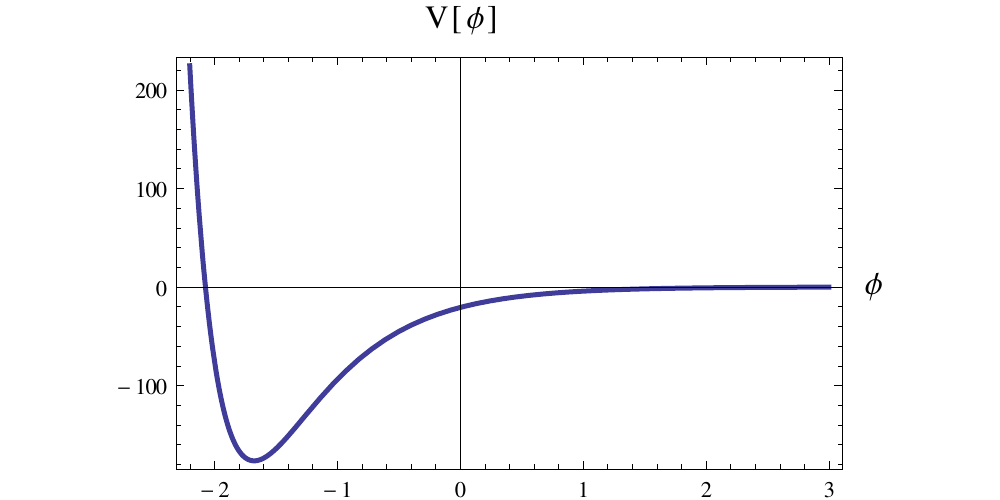}(b)\hspace{-1cm}\includegraphics[scale=0.8]{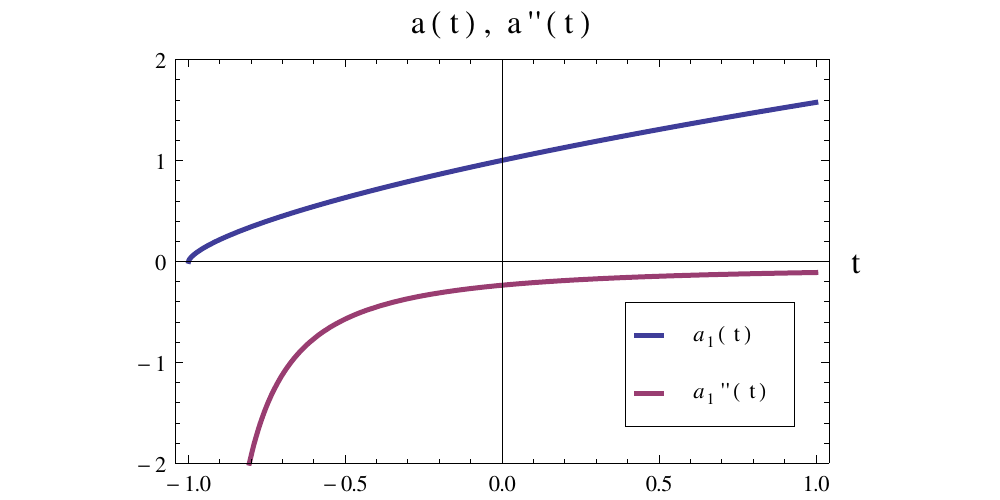}
\caption{Evolution of the potential $V(\varphi)$ versus $\varphi$. The fact of $V(\varphi)<0$ yields an inconsistent model with respect to 
observational data.}
\label{fig3}
\end{figure}
\begin{figure}[b]
\centering
\includegraphics{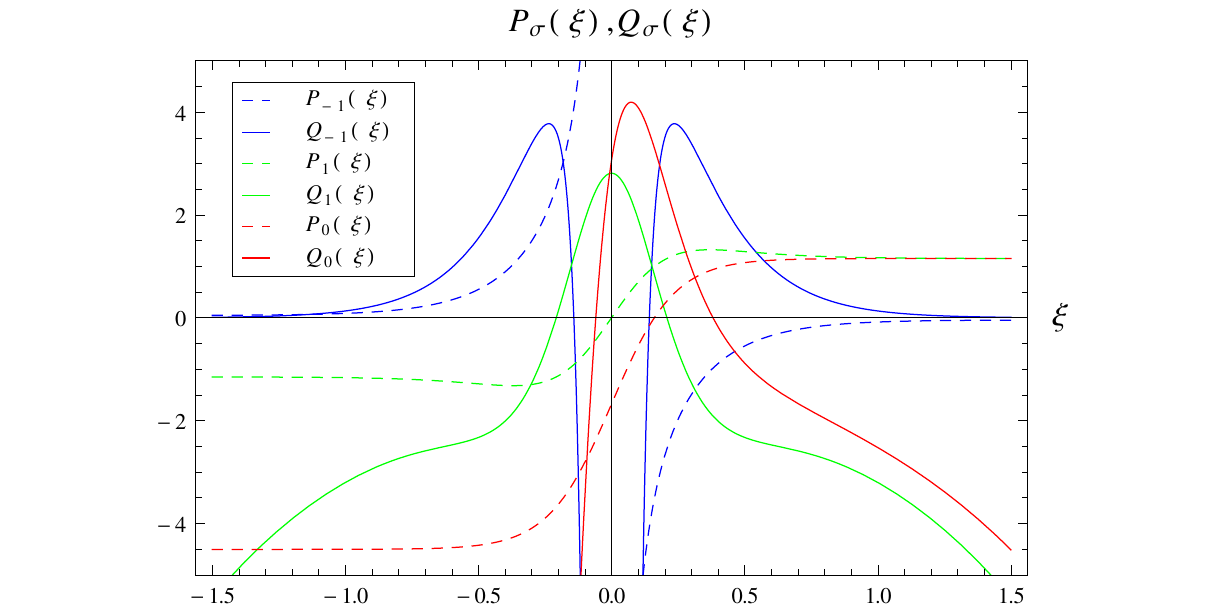}
\caption{The conditions \eqref{2.11}, where, again, the same convention for lines type and colors has been taken. Note that hyperbolic case also presents 
singular behavior around $\xi=0$.}\label{fig4}
\end{figure}
\paragraph{Spherical Case:} For this case, in figure \ref{fig3} (a), we can see that one big problem is found: the potential $V(\varphi)<0$ for 
most of $\varphi$ values. This yields whether the dark energy density can be negative or  the parameter state equation has positive values. Although 
this problem can be avoided by suitable setting of free constants, the description is just improved a little and it is not possible solve any 
cosmological problem under this scenario, as we can see in figure \ref{fig3} (b) where the $4$-dim acceleration is always negative.
%
%
\subsection{SM2\boldmath{$\bot$}SM2}
\begin{figure}[t]
\centering
(a)\hspace{-1cm}\includegraphics[scale=0.8]{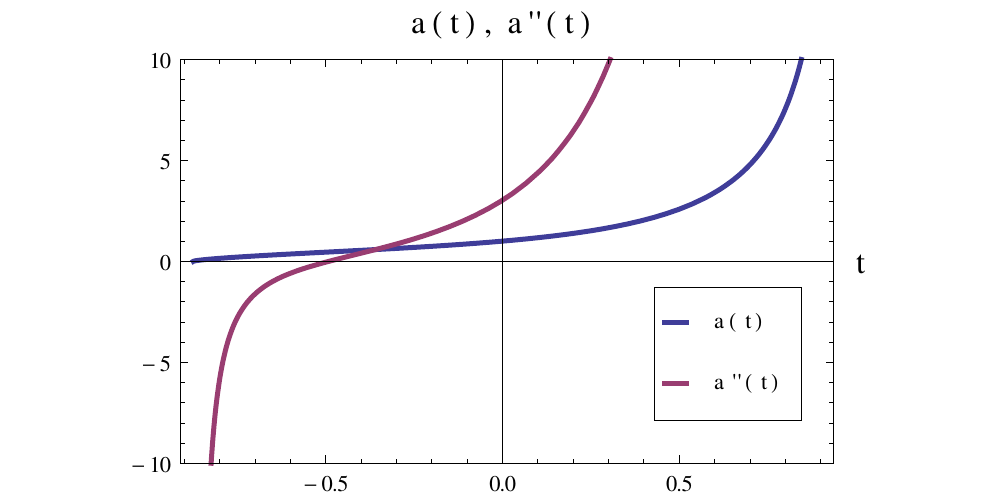}(b)\hspace{-1cm}\includegraphics[scale=0.8]{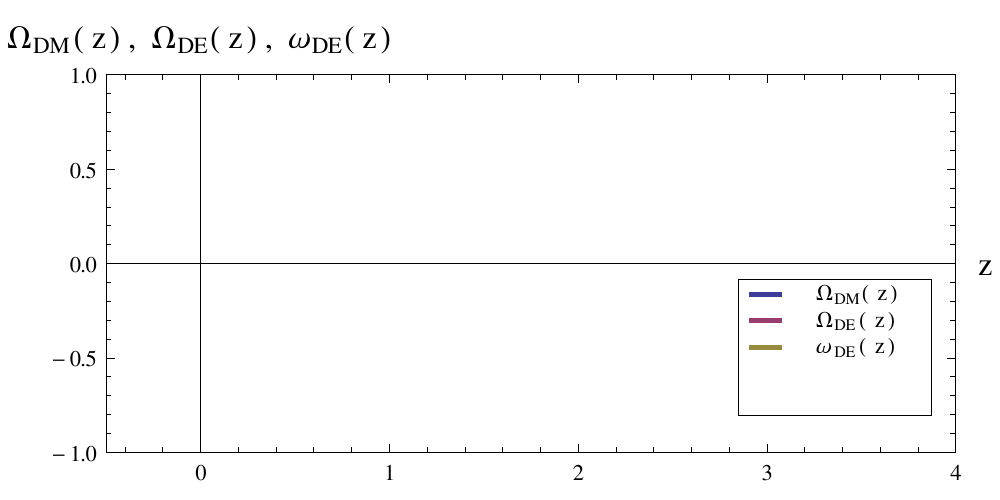}
\caption{Again, we have plotted the same functions like in figure \ref{fig2}, but this time for hyperbolic internal space 
in SM2$\bot$SM2 intersection.}\label{fig5}
\end{figure}
\begin{figure}[b]
\centering
(a)\hspace{-1cm}\includegraphics[scale=0.8]{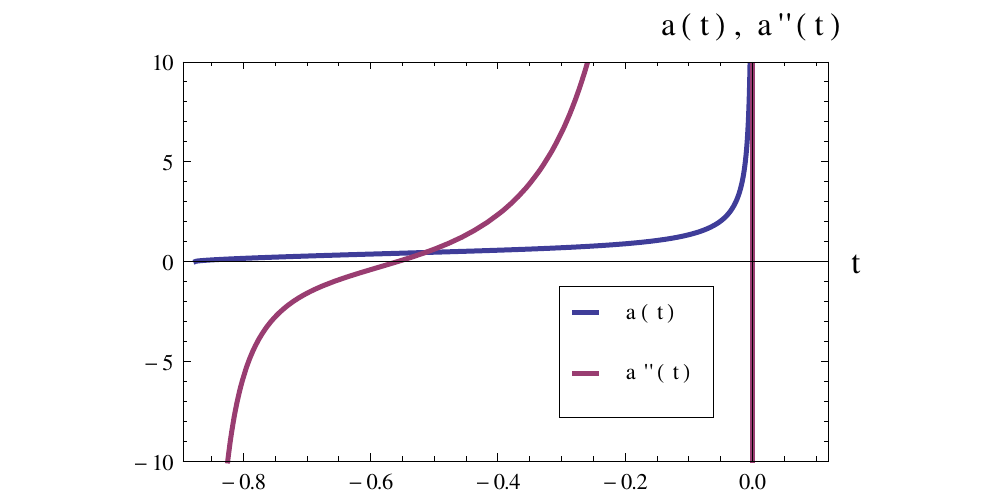}(b)\hspace{-1cm}\includegraphics[scale=0.8]{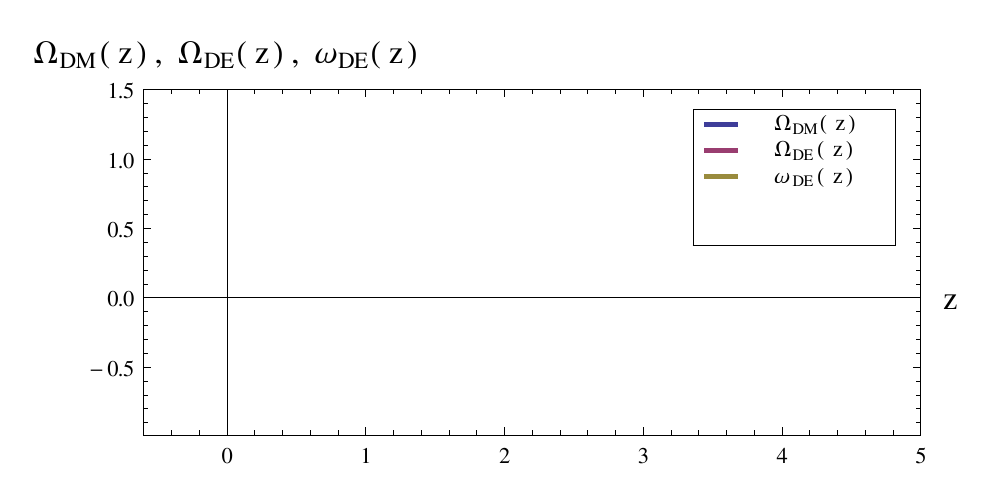}
\caption{Again, we have plotted the same functions like in figure \ref{fig2}, for plane internal space 
in SM2$\bot$SM2 intersection.}\label{fig6}
\end{figure}
\begin{figure}[t]
\centering
(a)\hspace{-1cm}\includegraphics[scale=0.8]{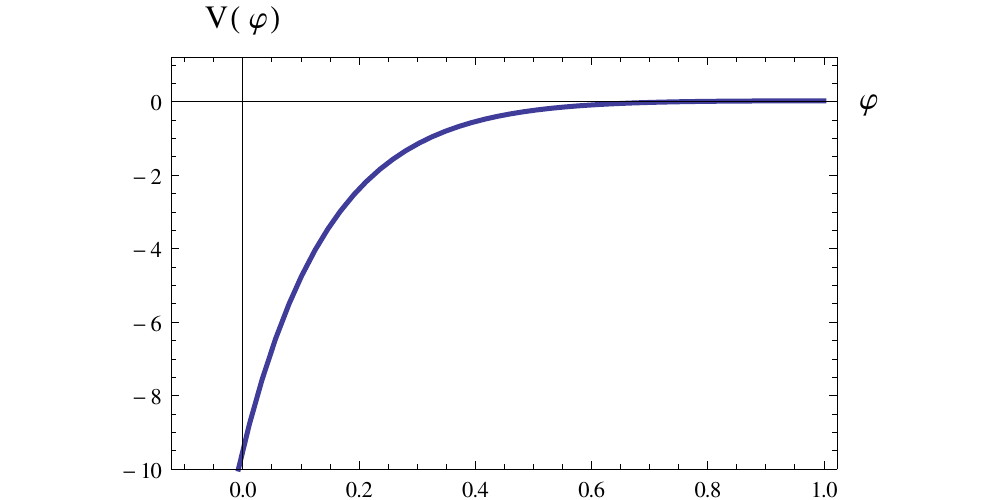}(b)\hspace{-1cm}\includegraphics[scale=0.8]{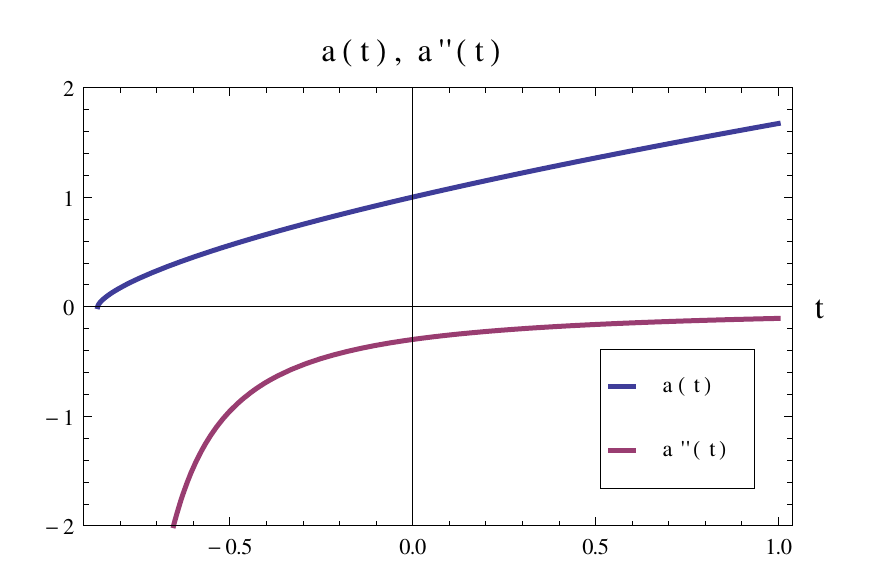}
\caption{Here, we have plotted the potential $V(\varphi)$ as in \ref{fig3}, for spherical internal space 
in SM2$\bot$SM2 intersection. As before, it is negative for most of $\varphi$ values.}\label{fig7}
\end{figure}

For this case, in addition to considerations in section \ref{sec2.2.2}, let us fix the $q$ and $m$ numerical values for H function in \eqref{2.16}, 
in 1 and 3, respectively, as we did for SM2 case. This is done with the purpose of analyzing the resulting scenario for two intersecting SM2-branes like was treated in previous section. Under these considerations, in figure \ref{fig4}, the behavior of $d$-dim scale factors is showed. We can see that is possible to get positive acceleration for certain intervals of time for all $\sigma$ values. However, not all of them works once you add CDM 
as we shall see.
\paragraph{Hyperbolic Case:} In figure \ref{fig5} (a), we can see $4$-dim scale factor $a(t)$ behavior, which evolves from the big bang at $t\approx-1$ until its present value $a(0)\approx1$ and continues speeding up forward the future. After approximately half the universe life time, the 
acceleration remains positive and increases its value continuously. This $4$-dim scenario corresponds quite well with the current evidence, which tells us that the universe could be eventually formed by the intersection of two SM-branes. \newline
Additionally, the fact $\omega_{DE}<-0.74$ as is required, makes us think about the possibility to improve our results making different considerations. It's remarkable and pretty unexpected how the dark energy dominates at the initial stages but later decreases relatively fast. After both energy contributions have the same value for $z\approx1$, dark matter begins to dominate. Despite all of this, if we look at numerical values and evolution of dark matter and dark energy relative densities $\Omega_{DM}$, $\Omega_{DE}$, as well as state equation parameter $\omega_{DE}$, the model fails and does not correspond with the accepted standard, i. e., the model for two intersecting SM2-branes under our conditions, works partially.
\paragraph{Plane Case:} Repeating the procedure for this geometry, we find that, as is showed in figure \ref{fig6}, neither the  $4$-dim scale factor $a(t)$ nor the numerical values for dark contributions correspond to what is expected for them. In figure \ref{fig6} (a), we see that despite we get positive acceleration $a(t)$ never reaches its current accepted value. 
In figure \ref{fig6} (b), we find also inconsistent behaviors for the plotted properties, e.g. $\omega_{DE}$ has value $-0.5$ at $z=0$.
\paragraph{Spherical Case:} Finally, as we expected, in figure \ref{fig7} (a), we find once again an inadequate behavior for the potential $V(\varphi)$, which eliminates any possibility as was discussed above. Despite this, in figure \ref{fig7} (b), $a(t)$ grows as is expected but its acceleration is always negative, emphasizing even more how much fails this model 
when internal space has spherical geometry.
%
%
\subsection{SM2\boldmath{$\bot$}SM5}
\begin{figure}[b]
\centering
\includegraphics{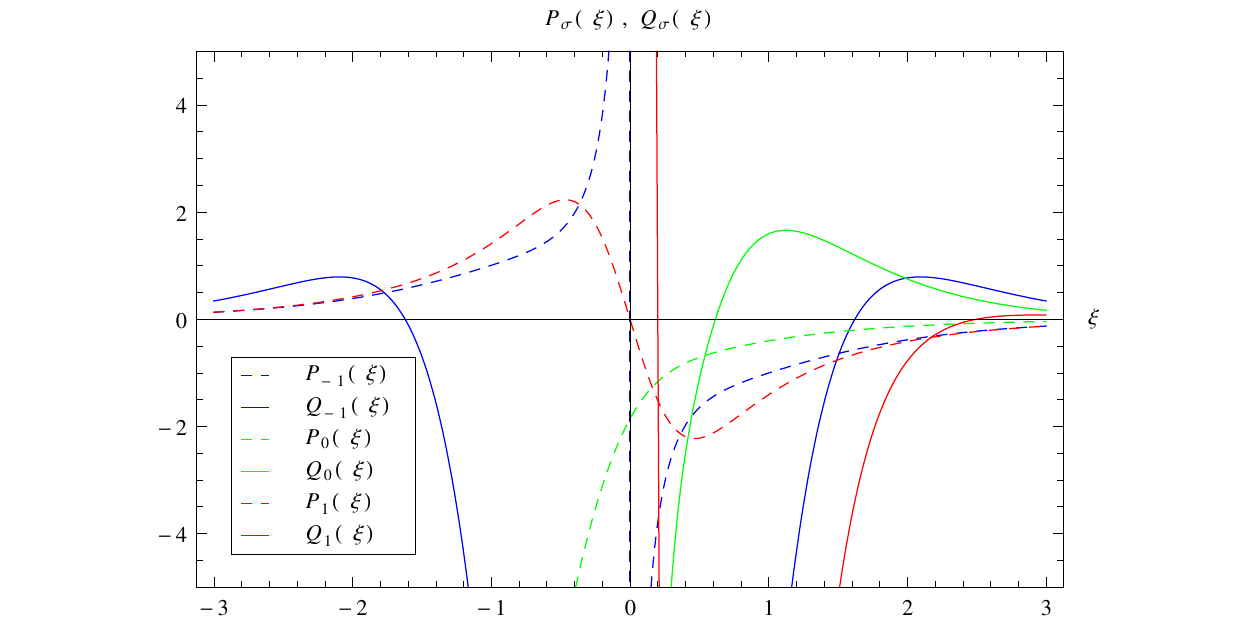}
\caption{The conditions \eqref{2.11}, where, once more, the same convention for lines type and colors has been taken. Positive acceleration is obtained 
for all cases in certain small intervals of time.}\label{fig8}
\end{figure}
\begin{figure}[t]
\centering
(a)\hspace{-1cm}\includegraphics[scale=0.8]{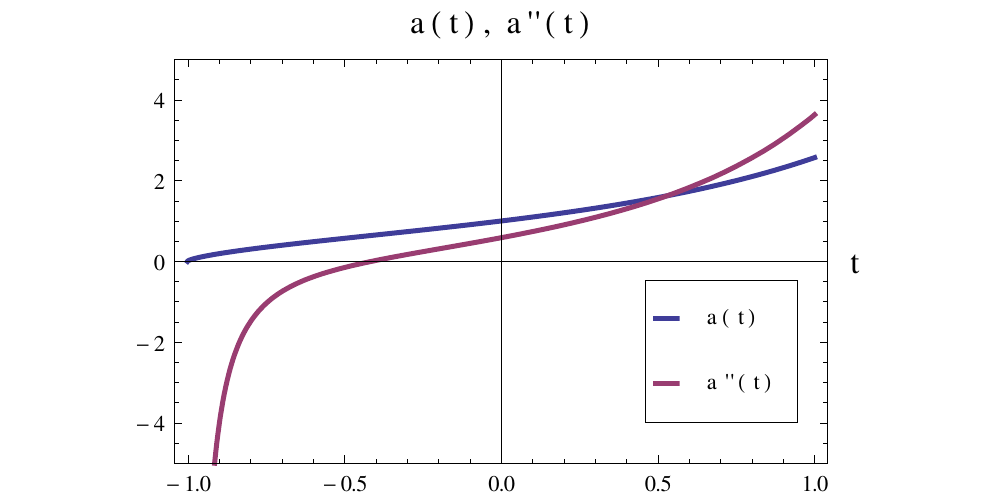}(b)\hspace{-1cm}\includegraphics[scale=0.8]{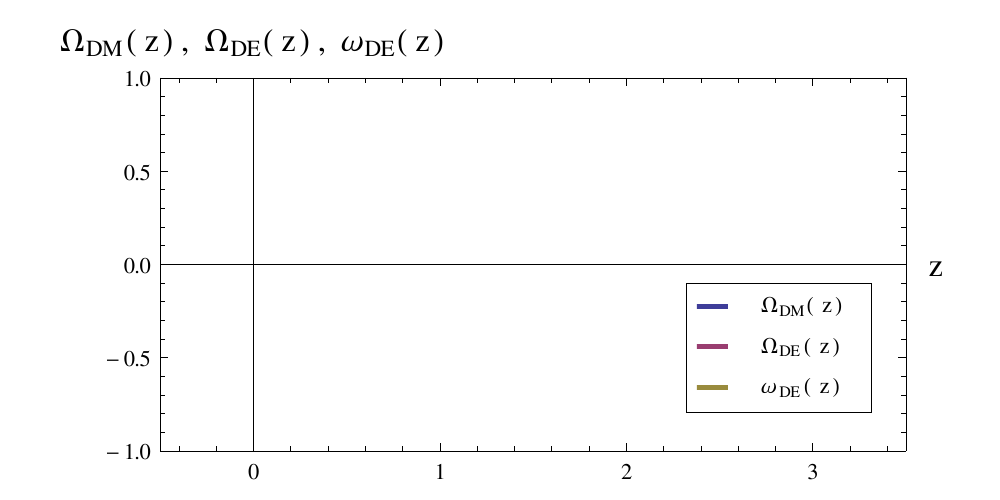}
\caption{Again, we have plotted the same functions like in figure \ref{fig2}, for hyperbolic internal space 
in SM2$\bot$SM5 intersection.}\label{fig9}
\end{figure}
\begin{figure}[b]
\centering
(a)\hspace{-1cm}\includegraphics[scale=0.8]{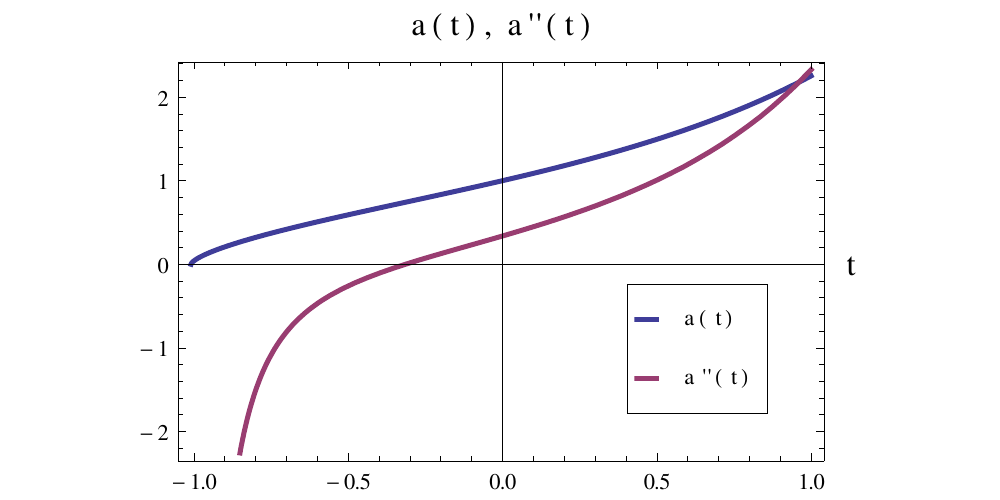}(b)\hspace{-1cm}\includegraphics[scale=0.8]{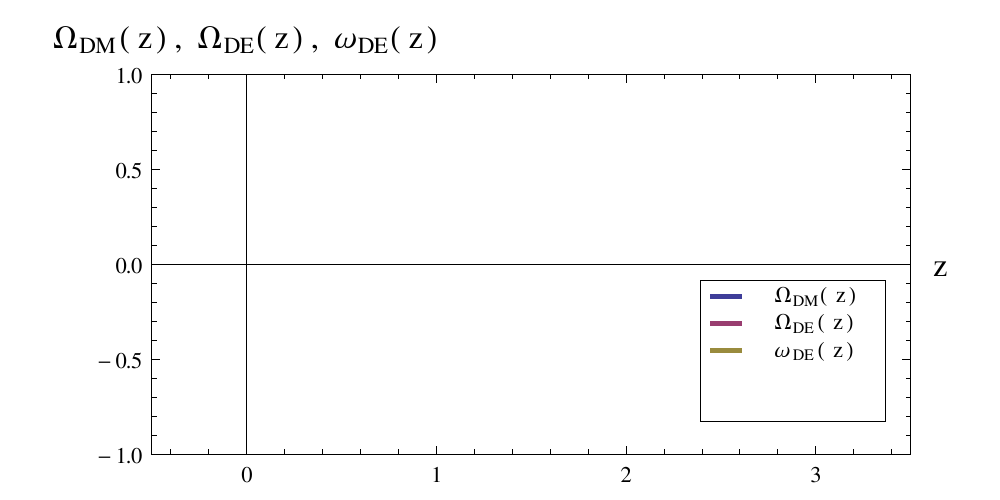}
\caption{Again, we have plotted the same functions like in figure \ref{fig2}, for plane internal space 
in SM2$\bot$SM5 intersection.}\label{fig10}
\end{figure}
\begin{figure}[t]
\centering
(a)\hspace{-1cm}\includegraphics[scale=0.8]{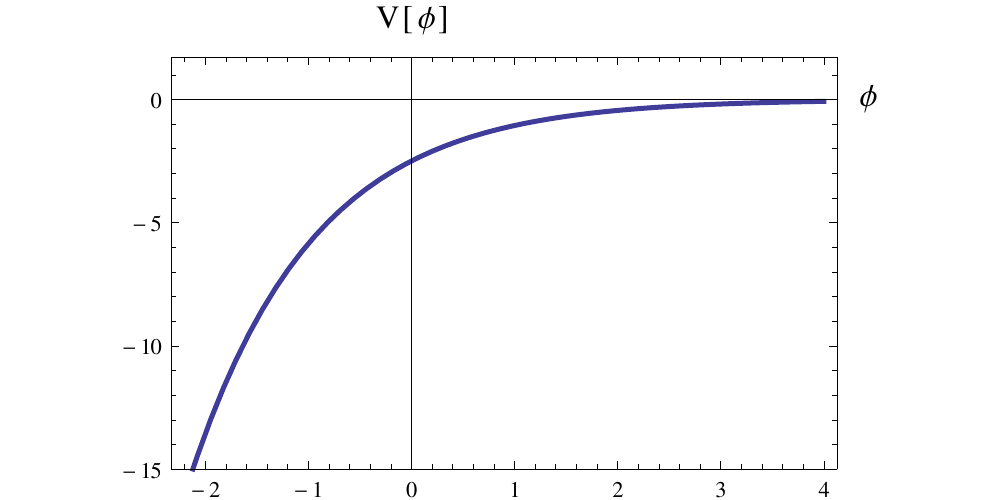}(b)\hspace{-1cm}\includegraphics[scale=0.8]{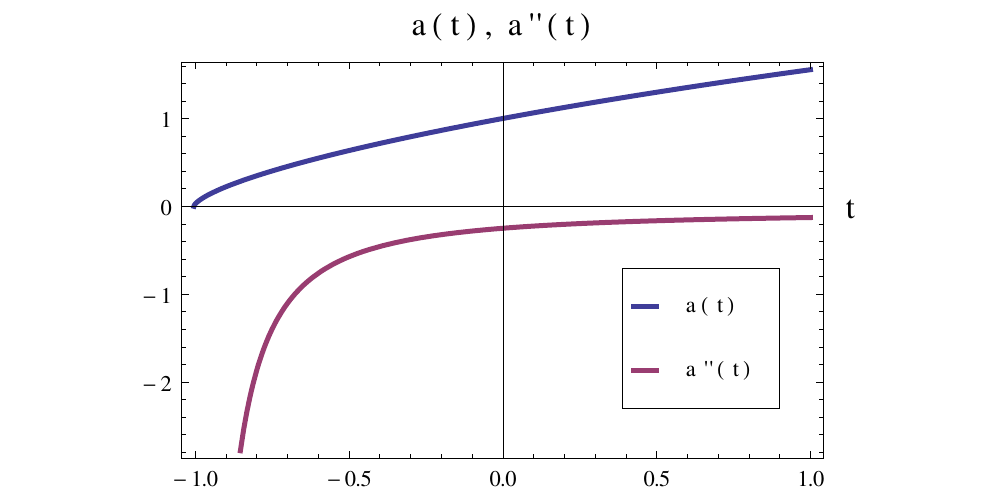}
\caption{Here, we have plotted the potential $V(\varphi)$ as in \ref{fig3}, for spherical internal space 
in SM2$\bot$SM5 intersection. As before, is negative for most of $\varphi$ values.}\label{fig11}
\end{figure}

%
%
In this case, let us consider the positive numbers $M_i$ in $H_i$ functions being equal, i. e., both branes having the same time-scaling value. We shall not proceed in the same way with $q_i$ numbers (charges), because it is interesting to investigate what would yield if proposing another relationship between them. 
Here, for the sake of simplicity, let us consider a linear dependence, i.e., $q_2=\alpha q_1$, 
where $\alpha$ is a fixed parameter. By this process, we find as final result that behavior of $d$-dim scale factors 
and their derivatives, only vary slightly under this assumption. Therefore, we choose the values of $\alpha$ and $q_1$ so that we could get the best possible description. In figure \ref{fig8}, as above, the evolution of $P(\xi)$ and 
$Q(\xi)$ is showed for all $\sigma$ values. As is possible to get positive acceleration, let us analyze how the $4$-dim model evolves 
including CDM for each geometry, as follows.
\paragraph{Hyperbolic Case:} Now we have added CDM, in figure \ref{fig9} it is interesting note in cosmic related parameters evolution for this geometry, that scale factor $a(t)$ grows as we expected, we get positive acceleration and the numerical current values for $\Omega_{DM}$, $\Omega_{DE}$ and $\omega_{DE}$, which are $\approx$0.32, $\approx$0.67 and $\approx$-0.89, respectively, are closer to accepted standard than before. Once again, dark energy component dominates at the ``initial stage'' ($z=0$ or $a=1$), but after both dark components reach the same value ($z\approx0.22$), dark matter becomes the cosmological dominant component at the infinite future.
\paragraph{Plane Case:} As in all the above models for this geometry, although it is possible to describe acceleration, the numerical values for ``dark parameters'' are too far away from reality. This can be seen in figure \ref{fig10}.
\paragraph{Spherical Case:} Finally, as we expected, spherical geometry does not work, because again the potential is a negative valued function. The $4$-dim acceleration is also negative. This can be seen in figure \ref{fig11}.
%
%
\section{Conclusions}
Initially, we have reviewed the study of cosmic scenarios from a model in M-theory when the universe is completely dark energy dominated in $d$-dim via scale factors evolution and when the cold dark matter is added phenomenologically into the $4$-dim cosmic evolution equations. Only the dark energy component is deduced from $d$-dim formulation and is represented by a scalar field which characterizes the internal space volume named radionic field. It has been exposed the main reasons why the plane and spherical cases do not work like hyperbolic geometry does in one single SM2 scenario, analyzing the numerical values of $\Omega_{DM}$, $\Omega_{DE}$ and $\omega_{DE}$ for each case as well as the scalar potential behavior $V(\varphi)$. Under special and approximate conditions, we have studied those properties applying the same method for intersecting branes SM2$\bot$SM2 and SM2$\bot$SM5 cases, finding that always the hyperbolic geometry describes better the current cosmological state. It is remarkable and new the fact that we can reproduce quite well the standard values for the studied cosmic parameters in several cases, mainly for SM2$\bot$SM5 with hyperbolic geometry. It seems attractive to give 
continuity to this kind of developments, maybe our universe can be formed by one or several SM-branes intersection.
%
%
\acknowledgments{Agudelo and Arcos appreciate the support given by Colciencias and Universidad Tecnológica de Pereira, in the framework of the Project 566 - 2012. }
%
%
\appendix 
%
%
%
%
\section{Harmonic Rules for S-branes in M-theory}\label{A}
\small{
We start form the general solution for a number $m$ of $Sq-Branes$ that intersect orthogonally, proposed by N. Ohta above mentioned in \cite{Ohta:2003-2}, this is
\begin{eqnarray}\label{A.0} \nonumber
ds_{d}^{2}&=&\prod_{A}\left[\cosh\widetilde{c_A}(\xi-\xi_A)\right]^{2\frac{q_{A}+1}{\Delta_A}}
\Bigl[e^{2ng(\xi)+2c_{0}\xi+c'_{0}}
\cdot\left(-d\xi^{2}+e^{-2(n-1)g(\xi)}d\Sigma^{2}_{n,\sigma}\right)\\
&&+\sum_{\alpha=1}^{p}\prod_{A}\left[\cosh\widetilde{c}_A(\xi-\xi_A)\right]^{-2\frac{\gamma_{A}^{(\alpha)}}{\Delta_A}}e^{2\widetilde
{c_{\alpha}}\xi+2c'_{\alpha}}dy_{\alpha}^{2}\Bigr],
\end{eqnarray}
where the involved functions and constants are given by
\begin{eqnarray}\label{A.1}
    \nonumber c_0&=&\sum_{A}\frac{q_{A}+1}{\Delta_A}\tilde{c_A}-\frac{1}{n-1}\sum_{\alpha=1}^{p}c_{\alpha},\\
   \nonumber c'_0&=&-\frac{1}{n-1}\sum_{\alpha=1}^{p}c'_{\alpha},\\
    \tilde{c}_{\alpha}&=&c_{\alpha}-\sum_{A}\frac{\delta_{A}^{(\alpha)}+1}{\Delta_A}\tilde{c_A},\\
   \nonumber \tilde{c}_{\phi}&=&c_{\phi}+\sum_{A}\frac{(d-2)\epsilon_{A}a_{A}}{\Delta_A}\tilde{c_A},
\end{eqnarray}
and
\begin{equation}
  \nonumber  \Delta_{A}=(q_{A}+1)(d-q_{A}-3)+\frac{1}{2}{a_{A}}^2(d-2),
\end{equation}
\begin{equation}
\gamma_{A}^{(\alpha)}=\left\{
           \begin{array}{ll}
             d-2, & \hbox{for $y_{\alpha}\epsilon q_{A}$;}  \\
             0, & \hbox{otherwise.} \\
           \end{array}
         \right.
\end{equation}
We also have,
\begin{equation}\label{A.3}
 g(t)=\left\{
           \begin{array}{ll}
             \frac{1}{n-1}\ln\frac{\beta}{\sinh[(n-1)\beta|\xi-{\xi}_{0}|]}, & \hbox{$\sigma=-1$;} \\
             \pm\beta(\xi-{\xi}_{0}), & \hbox{$\sigma=0$;} \\
             \frac{1}{n-1}\ln\frac{\beta}{\cosh[(n-1)\beta|\xi-{\xi}_{0}|]}, & \hbox{$\sigma=+1$.}
           \end{array}
         \right.,\\ 
\end{equation}
with
\begin{equation}\label{A.4}
    \beta=\pm\sqrt{\frac{1}{n(n-1)}\left[\frac{1}{n-1}\left(\sum_{\alpha=1}^{p}c_{\alpha}\right)^{2}+
    \sum_{\alpha=1}^{p}c_{\alpha}^{2}+\frac{1}{2}c_{\phi}^{2}\right]}
\end{equation}
Let us make the next definition:
\begin{equation}\label{A.5}
   \Upsilon_{A}^{(\alpha)}=\left\{
           \begin{array}{ll}
             \frac{-1}{q_{A}+1}, & \hbox{for $y_{\alpha}\in q_{A}-brane$;}  \\
             \frac{1}{d-q_{A}-3}, & \hbox{otherwise.} \\
           \end{array}
         \right.
\end{equation}
We shall build a more particular solution which is valid only in M-Theory framework $(d=11, 
a_{A}=0)$. In this sense, we start by restrict some integration constants in order to satisfy 
the condition:
\begin{equation}\label{A.6}
   c_{\alpha}=-\sum_{A}\Upsilon_{A}^{(\alpha)}\tilde{c_{A}}
\end{equation}
Is easy to note from \eqref{A.5}, for each $q_{A}-brane$ is true that:
\begin{equation}\label{A.7}
    \sum_{\alpha=1}^{p}\Upsilon_{A}^{(\alpha)}=\frac{1-n}{d-q_{A}-3}
\end{equation}
Replacing \eqref{A.6} (taking into account \eqref{A.7}), in equations \eqref{A.1}, we get:
\begin{equation}\label{A.8}
    \tilde{c_{\alpha}}=c_{0}=0
\end{equation}
Consequently, in the M-Theory framework and with the condition \eqref{A.6}, we can rewrite the 
metric given by \eqref{A.0} as
\begin{eqnarray}\label{A.9}\nonumber
ds_{d}^{2}&=&\prod_{A}\left[\cosh^{2}\tilde{c_A}(\xi-\xi_A)\right]^{\frac{1}{d-q_{A}-3}}
\Bigl\{e^{2ng(\xi)}e^{2c'_{0}}\left[-d\xi^{2}+e^{2(1-n)g(\xi)}d\Sigma^{2}_{n,\sigma}\right]\\
&&+\sum_{\alpha=1}^{p}\prod_{A}\left[\cosh^{2}\tilde{c_A}(\xi-\xi_A)\right]^{\Upsilon_{A}^{(\alpha)}}e^{2c'_{\alpha}}dy_{\alpha}^{2}\Bigr\}
\end{eqnarray}
We rename the constants as follows
\begin{equation}\label{A.10}
e^{c'_{\alpha}}=\prod_{A=1}^{m}\left(\frac{Q_{A}}{\tilde{c_{A}}}\right)^{\Upsilon_{A}^{(\alpha)}}
\end{equation}
Using the expression for $c'_{0}$ given by \eqref{A.1}, with the above condition and taking into account \eqref{A.7}, one can show that
\begin{equation}\label{A.11}
e^{c'_{0}}=\prod_{A=1}^{m}\left(\frac{Q_{A}}{\tilde{c_{A}}}\right)^{\frac{1}{d-q_{A}-3}}
\end{equation}
Furthermore, using \eqref{A.10} and \eqref{A.11} we can rewrite \eqref{A.9} as
\begin{eqnarray}\label{A.12}\nonumber
ds_{d}^{2}&=&\prod_{A}\left[\left(\frac{Q_{A}}{\tilde{c_{A}}}\right)^{2}\cosh^{2}\tilde{c_A}(\xi-\xi_A)\right]^{\frac{1}{d-q_{A}-3}}
\Bigl\{e^{2ng(\xi)}\left[-d\xi^{2}+e^{2(1-n)g(\xi)}d\Sigma^{2}_{n,\sigma}\right]\\
&&+\sum_{\alpha=1}^{p}\prod_{A}\left[\left(\frac{Q_{A}}{\tilde{c_{A}}}\right)^{2}\cosh^{2}\tilde{c_A}(\xi-\xi_A)\right]^{\Upsilon_{A}^{(\alpha)}}dy_{\alpha}^{2}\Bigr\}
\end{eqnarray}
We also define the next function
\begin{equation}\label{A.13}
G_{n,\sigma}=e^{2(1-n)g(\xi)}
\end{equation}
therefore
\begin{equation}\label{A.14}
e^{2kg(\xi)}=G_{n,\sigma}^{-\frac{n}{n-1}}.
\end{equation}
If we introduce the next constant
\begin{equation}\label{A.15}
 M=\sqrt{\frac{2}{n-1}\left(\sum_{\alpha=1}^{p}c_{\alpha}\right)^{2}+2\sum_{\alpha=1}^{p}c_{\alpha}^{2}+c_{\phi}^{2}}
\end{equation}
from equation \eqref{A.4}, we have
\begin{equation}\label{A.16}
   \beta=\pm\frac{M}{\sqrt{2n(n-1)}}.
\end{equation}
Replacing \eqref{A.16} in \eqref{A.3} and choosing $\xi_{1}=0$, which is possible by a simple shift in time, it follows that
\begin{equation}\label{A.17}
    g(\xi)=\left\{
           \begin{array}{ll}
             \frac{\ln\left(\frac{\sqrt{2n(n-1)}}{M}\sinh[\sqrt{\frac{n-1}{2n}}M|\xi|]\right)}{1-n}, & \hbox{$\sigma=-1$;} \\
             \frac{M}{\sqrt{2n(n-1)}}t, & \hbox{$\sigma=0$;} \\
             \frac{\ln\left(\frac{\sqrt{2n(n-1)}}{M}\cosh[\sqrt{\frac{n-1}{2n}}M \xi]\right)}{1-n}, & \hbox{$\sigma=+1$}
           \end{array}
         \right.
\end{equation}
where, we have taken
\begin{equation}\label{A.18}
   \beta=\left\{
           \begin{array}{ll}
             \frac{M}{\sqrt{2n(n-1)}}, & \hbox{for $\xi\geq0$;}  \\ 
              -\frac{M}{\sqrt{2n(n-1)}}, & \hbox{for $\xi<0$.} \\
           \end{array}
         \right.
\end{equation}
Substituting \eqref{A.17} in \eqref{A.13}, we get
\begin{equation}\label{A.19}
    G_{n,\sigma}=\left\{
           \begin{array}{ll}
             \frac{2n(n-1)}{M^{2}}\sinh^{2}[\sqrt{\frac{n-1}{2n}}M|\xi|], & \hbox{$\sigma=-1$;} \\
             \frac{2n(n-1)}{M^{2}}\cosh^{2}[\sqrt{\frac{n-1}{2n}}M\xi], & \hbox{$\sigma=+1$;} \\
            e^{2M\sqrt{\frac{n-1}{2n}}}, & \hbox{$\sigma=0$.}
           \end{array}
         \right.
\end{equation}
Now, defining the functions
\begin{eqnarray}\label{A.20}
H_{A}&=&\left[\left(\frac{Q_{A}}{\tilde{c_{A}}}\right)^{2}\cosh^{2}\tilde{c_A}(\xi-\xi_A)\right]^{\frac{1}{d-q_{A}-3}}\\
\label{A.21}
\tilde{H_{A}}&=&\left[\left(\frac{Q_{A}}{\tilde{c_{A}}}\right)^{2}\cosh^{2}\tilde{c_A}(\xi-\xi_A)\right]^{\frac{-1}{q_{A}+1}}\\
\label{A.22}
H_{A}^{(\alpha)}&=&\left\{
           \begin{array}{ll}
           H_{A} , & \hbox{for $y_{\alpha}\in q_{A}-brane$;}  \\
              \tilde{H_{A}}, & \hbox{otherwise;} \\
           \end{array}
         \right.
\end{eqnarray}
then using \eqref{A.22} and \eqref{A.19} (taking into account \eqref{A.13} and \eqref{A.14}), we can rewrite 
\eqref{A.12} as
\begin{equation}\label{A.23}
ds_{d}^{2}=\prod_{A}H_{A}G_{n,\sigma}^{-\frac{n}{n-1}}\Bigl[-d\xi^{2}+G_{n,\sigma}d\Sigma^{2}_{n,\sigma}\Bigr]
+\sum_{\alpha=1}^{p}\prod_{A}H_{A}^{(\alpha)}dy_{\alpha}^{2}
\end{equation}
Therefore, we can conclude that the metric \eqref{2.3} is totally equivalent to \eqref{A.23}, if the conditions 
\eqref{A.6}, \eqref{A.10} and \eqref{A.15}, are fulfilled,  into the M-Theory framework $(d=11, a_{A}=0)$.
%
%
\section{Dimensional Reduction}\label{B}
This process consist in obtaining $4$-dim gravity coupled to scalar field from $d$-dim. We start from
\begin{equation}\label{B.1}
\mathcal{L}=\frac{1}{2}C\sqrt{-g_{d}}R_{d},
\end{equation}
where $C=1/8\pi G_{d}$ and we are considering  $n=d-4$. The $n$-dim space is compactified to a small region, letting the $4$-dim part extended infinitely in space-time. The corresponding field equations to (\ref{B.1}) are
\begin{equation}\label{B.2}
 R_{AB}-\frac{1}{2}R g_{AB}=0,\quad A,B=0,1,\dots,d-1,
\end{equation}
where $g_{AB}$ is the metric of $d$-dim total space-time. The next step is to interpret the $d$-dim  metric tensor $g_{AB}$, as being composed of one observable (external) and other non-observable (internal) parts. Assuming that total $d$-dim space-time manifold $N$ is formed by he product
\begin{equation}\label{B.3}
  N_{d}=L_{4}\times M_{n}\rightarrow
  g_{AB}=\left(
           \begin{array}{cc}
             g_{\mu\nu}(x^{\rho}) & 0 \\
             0 & {g}_{ik}(x^{c}) \\
           \end{array}
         \right),
\end{equation}
where $\mu, \nu=0,1,2,3$ e $i, k=4,...,d-1$, the metric ${g}_{ik}$ is associated to internal compact space. With the aim to attribute dynamical properties to that space, we are going to consider
\begin{equation}\label{B.4}
 g_{AB}=\left(
           \begin{array}{cc}
             e^{-n\psi(x^\mu)}\bar{g}_{\mu\nu}(x^{\rho}) & 0 \\
             0 & e^{2\psi(x^{\mu})}\tilde{g}_{ik}(x^{j}) \\
           \end{array}
         \right),
\end{equation}
such that
\begin{equation}\label{B.5}
 g_{\mu\nu}=e^{-n\psi(x^\mu)}\bar{g}_{\mu\nu} \quad and \quad g_{ik}=e^{2\psi(x^{\mu})}\tilde{g}_{ik}(x^{j}).
\end{equation}
On the other hand, the Christoffel symbols are defined as usual
\begin{equation}\label{B.6}
\Gamma\indices{^{A}_{BC}}=\frac{1}{2}g^{AD}\left(\partial_{B}g_{DC}+\partial_{C}g_{DB}-\partial_{D}g_{BC}\right).
\end{equation}
Computing the non-zero symbols, we have
\begin{eqnarray}\label{B.7}
 \Gamma\indices{^{\lambda}_{\mu\nu}}=\Gamma\indices{^{\lambda}_{\nu\mu}}, \quad
 \Gamma\indices{^{\mu}_{ij}}=\Gamma\indices{^{\mu}_{ji}}=-\frac{1}{2}g^{\mu\nu}\left(\partial_{\nu}g_{ij}\right)\nonumber\\
\Gamma\indices{^{i}_{\mu j}}=\Gamma\indices{^{i}_{j\mu}}=\frac{1}{2}g^{ik}\left(\partial_{\mu}g_{kj}\right), \quad
 \Gamma\indices{^{i}_{jk}}=\Gamma\indices{^{i}_{kj}}.
\end{eqnarray}
However, they are composed
\footnote{Here, according to \eqref{B.4} $\Omega=e^{-n\psi(x^\mu)}$}
\begin{equation}
 \Gamma\indices{^{\lambda}_{\mu\nu}}=\bar{\Gamma}\indices{^{\lambda}_{\mu\nu}}+
 \Omega^{-1}\Bigl[\delta\indices{^{\lambda}_{\nu}}\left(\partial_{\mu}\Omega\right)
 +\delta\indices{^{\lambda}_{\mu}}\left(\partial_{\nu}\Omega\right)-
 (\partial^{\lambda}\Omega)\bar{g}_{\mu\nu}\Bigr],
\end{equation}
\begin{equation}
\Gamma\indices{^{i}_{kj}}=\tilde{\Gamma}\indices{^{i}_{kj}}.\label{B.8}
\end{equation}
and, specifically, we obtain
\begin{eqnarray}\label{B.9}
 \Gamma\indices{^{\rho}_{ik}}&=&\Gamma\indices{^{\rho}_{ki}}=-e^{(n+2)\psi}\bar{g}^{\rho\nu}\left(\partial_{\nu}\psi\right)\tilde{g}_{ik}\nonumber\\
 \Gamma\indices{^{i}_{\mu j}}&=&\Gamma\indices{^{i}_{j\mu}}=\delta\indices{^{i}_{j}}\left(\partial_{\mu}\psi\right).
\end{eqnarray}
We can now calculate
\begin{equation}\label{B.10}
 R_{AB}=R_{\mu\nu}+R_{ik},
\end{equation}
remembering that
\begin{equation}\label{B.11}
 R_{AB}=\partial_{C}\Gamma\indices{^C_{AB}}+\Gamma\indices{^C_{CD}}\Gamma\indices{^D_{AB}}-
 (A\leftrightarrow B).
\end{equation}
Thus, the $4$-dim and $n$-dim  components of Ricci tensor are given by
 \begin{eqnarray}\label{B.12}
   R_{\mu\nu}&=&\bar{R}_{\mu\nu}+\frac{n}{2}\square\psi\bar{g}_{\mu\nu}-\frac{n(n+2)}{2}(\partial_{\mu}\psi)(\partial_{\nu}\psi),\\
\label{B.13}
  R_{ik}&=&\tilde{R}_{ik}-e^{(n+2)\psi}\square\psi\tilde{g}_{ik},\\
\label{B.14}
  R_{i\mu}&=& R_{\mu i}=0,
\end{eqnarray}
with 
\begin{equation}\label{B.15}
  \square\psi=g^{\rho\lambda}\triangledown_{\lambda}\triangledown_{\rho}(\psi).
\end{equation}
Finally, the scalar of curvature is
\begin{equation}\label{B.16}
R=e^{n\psi}\bar{R}+e^{-2\psi}\tilde{R}-\frac{n}{2}(n+2)\left(\bar{\triangledown}\psi\right)^{2}+ne^{n\psi}\square\psi,
\end{equation}
where 
\begin{equation}\label{B.17}
\left(\bar{\triangledown}\psi\right)^{2}=\bar{g}_{\mu\nu}(\partial_{\mu}\psi)(\partial_{\nu}\psi).
\end{equation}
Here, $\bar{R}$ and $\tilde{R}$, represent the scalars of curvature $4$-dim and $n$-dim, respectively. In order to relate those quantities, we take the ($\mu\nu$) component in \eqref{B.2}, such that
\begin{equation}\label{B.18}
\bar{R}_{\mu\nu}-\frac{1}{2}\bar{g}_{\mu\nu}\bar{R}=\frac{n(n+2)}{2}\Bigl[(\partial_{\mu}\psi)(\partial_{\nu}\psi)-
\frac{1}{2}(\partial^{\rho}\psi)(\partial_{\rho}\psi)\bar{g}_{\mu\nu}\Bigr]-V(\psi)\bar{g}_{\mu\nu},
\end{equation}
with
\begin{equation}\label{B.19}
 V(\psi)=-\frac{\tilde{R}}{2}e^{-(n+2)\psi}.
\end{equation}
Now, multiplying by $e^{(n+2)\psi}\bar{g}^{\mu\nu}$, we have
\begin{equation}\label{B.20}
 2\tilde{R}=-e^{(n+2)\psi}\left[\bar{R}-\frac{n(n+2)}{2}\partial^{\rho}\psi\partial_{\rho}\psi\right].
\end{equation}
The L.H.S and R.H.S of this expression are only $x^{a}$  and $x^{\mu}$ dependent, respectively, therefore, both are equal to one constant, namely, $2K$. Thus, we get
\begin{equation}\label{B.21}
 \tilde{R}=K,
\end{equation}
\begin{equation}\label{B.22}
  -e^{(n+2)\psi}\left[\frac{\bar{R}}{2}-\frac{n(n+2)}{4}\partial^{\rho}\psi\partial_{\rho}\psi\right]=K,
\end{equation}
and also
\begin{equation}\label{B.23}
 \tilde{R}=\tilde{g}^{ik}\tilde{R}_{ik}=K \hspace{1cm} \rightarrow \hspace{1cm} \tilde{R}_{ik}=\frac{K}{n}\tilde{g}^{ik}.
\end{equation}
Using this last equation, we can re-write \eqref{B.16} as being
\begin{equation}\label{B.24}
R=e^{n\psi}\bar{R}+Ke^{-2\psi}-\frac{n(n+2)}{2}e^{n\psi}\left(\bar{\triangledown}\psi\right)^{2}+ne^{n\psi}\square\psi.
\end{equation}
On the $n$-dim manifold, it must be satisfied that
\begin{equation}\label{B.25}
 \tilde{R}_{ik}=\sigma(n-1)\tilde{g}_{ik},\quad \rightarrow \quad\tilde{R}=K=\sigma n(n-1)
\end{equation}
Hence, substituting \eqref{B.24} and \eqref{B.25} in \eqref{B.1} and building the corresponding $d$-dim action, we will have
\begin{equation}\label{B.26}
 S_{d}=\frac{C}{2}\int d^{d}x\sqrt{-g_{d}}\Bigl[e^{n\psi}\bar{R}+\sigma n(n-1)e^{-2\psi}
 -\frac{n(n+2)}{2}e^{n\psi}(\bar{\triangledown}\psi)^{2}+ne^{n\psi}\square\psi\Bigr].
\end{equation}
The $d$-dim determinant would be
\begin{equation}\label{B.27}
 g_{d}=det[e^{-n\psi}\bar{g}_{\mu\nu}]det[e^{2\psi}\tilde{g}_{ik}]=e^{-2n\psi}\bar{g}\tilde{g}.
\end{equation}
Substituting this result in the above expression and separating the integration volumes, we find that
\begin{equation}\label{B.28}
 S_{d}=\frac{C}{2}\int_{V_{\mu}}d^{4}x\sqrt{-\bar{g}}\Bigl[\bar{R}-\frac{n(n+2)}{2}(\bar{\triangledown}\psi)^{2}+
 \sigma n(n-1)e^{-(n+2)\psi}+
 n\square\psi\Bigr]\cdot\Bigl[\int_{V_{i}}d^{n}x\sqrt{\tilde{g}}\Bigr],
\end{equation}
with $V_\mu$ and $V_i$ being the $4$-dim and $n$-dim integration volumes, respectively. Like $\int_{V_{i}}{d^{n}x\sqrt{\tilde{g}}}=V_n$, i.e., the internal space volume and $G_{d}=V_{n}G_{4}$, the action becomes in
\begin{equation}\label{B.29}
 S_{4}=\frac{C'}{2}\int_{V_{\mu}}d^{4}x\sqrt{-\bar{g}}\Bigl[\bar{R}-\frac{n(n+2)}{2}(\bar{\triangledown}\psi)^{2}+
 \sigma n(n-1)e^{-(n+2)\psi}\Bigr]+\int_{V_{\mu}}{d^{4}x\sqrt{-\bar{g}}n\square\psi}.
\end{equation}
Analyzing the last term and applying the divergence theorem, we see that
\begin{equation}\label{B.30}
\int_{V_{\mu}}{d^{4}x\sqrt{-\bar{g}}\square\psi}=\int_{\partial V_{\mu}}{\sqrt{-\bar{g}}\bar{g}^{\mu\nu}(\partial_{\nu})dS_{\mu}}.
\end{equation}
Since $\delta\bar{g}^{\mu\nu}=0$ in $\partial V_{\mu}$, so then
\begin{equation}\label{B.31}
 \int_{V_{\mu}}{d^{4}x\sqrt{-\bar{g}}n\square\psi}=0.
\end{equation}
Thus, we get the result
\begin{equation}\label{B.32}
 S_{4}=\frac{C'}{2}\int_{V_{\mu}}{d^{4}x\sqrt{-\bar{g}}\left[\bar{R}-\frac{n(n+2)}{2}(\bar{\triangledown}\psi)^{2}-2V(\psi)\right]},
\end{equation}
with 
\begin{equation}\label{B.33}
 V(\psi)=-\sigma n(n-1)e^{-(n+2)\psi},
\end{equation}
and
\begin{equation}\label{B.34}
 C'=\frac{1}{8\pi G_{4}}.
\end{equation}
After this process, we can conclude two important aspects
\begin{itemize}
 \item The metric $\tilde{g}_{ab}$ of the internal space $M_{n}$ corresponds to a manifold with constant curvature scalar.
 \end{itemize}
 \begin{equation}
 \tilde{R}=K\Longrightarrow\tilde{R}_{ab}=\frac{K}{n}\tilde{g}_{ab}.\label{B.35} 
 \end{equation}
\begin{itemize}
 \item The scalar of curvature of the manifold $N_d$ only depends of the \emph{external} coordinates belonging to $L_4$.
\end{itemize}
\begin{equation}
  R=e^{n\psi}\bar{R}+K e^{-2\psi}-e^{n\psi}\frac{n(n+2)}{2}\left(\bar{\triangledown}\psi\right)^{2}+n e^{n\psi}\square\psi.\label{B.36}
 \end{equation}
The above procedure and expressions can be used to reduce a $d$-dim action of super-gravity to one scalar-tensor coupling gravity model in $4$-dim. 
%

}
%
%
%
%
\end{document}